\documentclass[12pt]{article}
\usepackage{amsfonts,amsmath}
\usepackage{amssymb}

\renewcommand{\theequation}{\thesection.\arabic{equation}}

\makeatletter \@addtoreset{equation}{section} \makeatother

{\vspace{3mm} }

\def\al{\alpha}

\def\*{\star}

\def\E2{\mathbf{E}}

\def\n{\overline{\nabla}}
\def\nn{\nabla}
\def\d{\overline{\square}}

\def\g{\overline{g}}

\newcommand{\be}{\begin{equation}}
\newcommand{\ee}{\end{equation}}
\newcommand{\bee}{\begin{eqnarray}}
\newcommand{\beee}{\begin{array}}
\newcommand{\eee}{\end{eqnarray}}
\newcommand{\eeee}{\end{array}}
\newcommand{\gpp}{\eta}%
\newcommand{\gn}{\nu}
\newcommand{\gm}{\mu}
\newcommand{\gc}{\chi}
\newcommand{\gx}{\xi}
\newcommand{\gr}{\rho}
\newcommand{\ga}{\alpha}
\newcommand{\gb}{\beta}
\newcommand{\gga}{\gamma}

\newcommand{\vf}{\varphi}
\newcommand{\gd}{\delta}
\newcommand{\gl}{\lambda}

\newcommand{\gs}{\sigma}

\newcommand{\q}{\,,\qquad}
\newcommand{\half}{\frac{1}{2}}

\newcommand{\p}{\partial}

\newcommand{\ff}{\frac}
\begin{document}
\begin{flushright}
FIAN/TD/14-2022\\
\end{flushright}

\vspace{0.5cm}
\begin{center}
{\large\bf Classical double copy and higher-spin fields}

\vspace{1 cm}

\textbf{V.~E.~Didenko${}^1$, N.~K.~Dosmanbetov${}^2$}\\

\vspace{1 cm}

\textbf{}\textbf{}\\
 \vspace{0.5cm}
 \textit{${}^1$ I.E. Tamm Department of Theoretical Physics,
Lebedev Physical Institute,}\\
 \textit{ Leninsky prospect 53, 119991, Moscow, Russia }\\

\vspace{0.7 cm}\textit{ ${}^2$ Department of General and Applied
Physics, Moscow Institute of
Physics and Technology,}\\
\textit{Institutskiy per. 7, Dolgoprudnyi, 141700 Moscow region,
Russia}

\par\end{center}

\begin{center}
\vspace{0.6cm}
didenko@lpi.ru, dosmanbetov.nk@phystech.edu\\

\par\end{center}

\vspace{0.4cm}

%
%


%

\begin{abstract}
\noindent Kerr-Schild double copy is shown to extend naturally to
all free symmetric gauge fields propagating on $(A)dS$ in any
dimension. Similarly to the standard lower-spin case, the
higher-spin multicopy comes along with the zeroth, single, and
double copies. The mass-like term of the Fronsdal spin $s$ field
equations fixed by gauge symmetry and the mass of the zeroth copy
both appear to be remarkably fine-tuned to fit the multicopy
pattern forming a spectrum organized by higher-spin symmetry. On
the black hole side this curious observation fills up the list of
miraculous properties of the Kerr solution.
\end{abstract}
\newpage
\section{Introduction}
Double copy relations between field theories originally based on
the observation from string theory \cite{Kawai:1985xq} have
evolved into relations for scattering amplitudes of gauge and
gravity theories \cite{Bern:2010ue}, \cite{Bern:2010yg}. In recent
years the field has become a subject of intense study
\cite{Adamo:2022dcm} (see also references therein). In particular,
the scattering amplitude philosophy has been extended to the level
of classical solutions revealing how certain gravity solutions
emerge as double copies of the gauge ones,
\cite{Monteiro:2014cda}.

The most well-known example is the Kerr black hole, which metric
casts into the Kerr-Schild form
\be\label{KS}
g_{\mu\nu}=\g_{\mu\nu}+M\vf k_{\mu}k_{\nu}\,.
\ee
Here, $\g_{\mu\nu}$ is the base metric that can be either
Minkowski or $(A)dS$, $M$ is a free parameter attributed to a
black hole mass, and $\vf$ and $k_{\mu}$ are the space-time
dependent scalar and vector, correspondingly. Remarkably, it then
turns out that vector potential $\vf_{\mu}=\vf k_{\mu}$ satisfies
the Maxwell equations, while $\vf$ satisfies the Klein-Gordon
equation\footnote{The scalar field equation is
$\overline\Box\vf=m^2_{\Lambda}\vf$, where the mass-like term is
given in terms of the cosmological constant.}. This makes gravity
perturbations $\vf_{\mu\nu}=\vf k_{\mu}k_{\nu}$ a 'square' of a
single copy $\vf_{\mu}$ up to a factor $\vf$ called the zeroth
copy. This fact was originally recognized using spinor language in
four dimensions \cite{Didenko:2008va}, \cite{Didenko:2009tc} in an
attempt to identify structures that may help to generalize a black
hole into a theory of interacting higher spins ($HS$)
\cite{Didenko:2009td} and in five dimensions in
\cite{Didenko:2011ir}. In the double copy literature it was
independently rediscovered in \cite{Monteiro:2014cda} for
asymptotically flat background.

From historical perspective, an early indication that black holes
should admit some doubling in terms of a spin $s=1$ field was
given in \cite{Walker:1970un}, \cite{Hughston:1972qf}, where it
was shown that for a black hole in particular
\be\label{wdc}
\textnormal{Weyl}\sim (\textnormal{Maxwell})^2\,.
\ee
This schematic relation is a consequence of the fact that the Kerr
solution is of Petrov type $D$ \cite{Petrov}. Property \eqref{wdc}
was later re-observed in \cite{Didenko:2008va},
\cite{Didenko:2011ir}, \cite{Luna:2018dpt} and dubbed in
\cite{Luna:2018dpt} the Weyl double copy.

The present literature on classical double copy is substantial
(see \cite{Ridgway:2015fdl}-\cite{Lescano:2022nhp} for an
incomplete list of references and \cite{Adamo:2022dcm} for more
therein) mostly confined to gauge/gravity cases. The results of
\cite{Didenko:2008va}-\cite{Didenko:2011ir} indicate, however,
that double copy can be extended beyond the realm of gauge/gravity
correspondence to include $HS$ fields $s>2$. Indeed, in
\cite{Didenko:2008va} it was shown that Kerr-Schild ansatz,
\eqref{KS} in four dimensions extends naturally to what can be
referred to as the multicopy
\be\label{anys}
\vf_{\mu_1\dots\mu_s}=\vf k_{\mu_1}\dots k_{\mu_s}\,,
\ee
which for $s=0$, $s=1$ and $s=2$ reproduces the known zeroth,
single and double copies respectively, while for $s>2$,
$\vf_{\mu_1\dots\mu_s}$ surprisingly satisfies the spin $s$
Fronsdal equations \cite{Fronsdal:1978rb}. The appearance of
massless fields of arbitrary spin as multicopies might have been
accidental thanks to the Penrose transform that generates the
whole tower in $d=4$ \cite{Didenko:2009td} (see also
\cite{White:2020sfn}). However, as shown in \cite{Didenko:2011ir},
the $HS$ Kerr-Schild and Weyl multicopies still exist in $d=5$ at
least at free level. For $HS$ interactions \cite{Vasiliev:1992av}
it has been shown recently how the Weyl multicopy shows up in
planar solutions at leading order \cite{Didenko:2021vui},
\cite{Didenko:2021vdb} that include a four dimensional black
brane. All that indicates that the classical $HS$ multicopy might
not be accidental being a phenomenon worth studying. Little is
known about what goes on at $d>5$ from that perspective.

In this paper, we address the question of whether the double copy
admits $HS$ generalization beyond lower dimensions where spinorial
isomorphisms may play a significant role. For that matter, we
revisit the $(A)dS$ Kerr solution in arbitrary dimensions of
\cite{Gibbons:2004js}.

Our main finding is easy to state. We show that the zeroth,
single, and double copies of the $AdS$-Kerr solution together
guarantee multicopy extension, \eqref{anys}, to all symmetric
massless fields of integer spins. The multicopy turns out to
satisfy the Fronsdal equations. A remarkable feature of the
observed multicopy is as follows. In order to be consistent with
the mass-like term of the spin $s\geq 1$ Fronsdal equations
\cite{Metsaev:1997nj},
\be\label{spect}
m^2_{s} = - \lambda((s-2)(d+s-3)-s)
\ee
the 'mass' of the zeroth copy of the $AdS$-Kerr solution should be
equal to
\be\label{m0}
m^2_0=2\gl (d-3)\,,
\ee
where $\gl$ is the cosmological constant. This turns out to be
exactly the case with Kerr. The spectrum of fields where the
multicopy is realized therefore consists of symmetric fields in
which 'masses' are given by \eqref{spect} for all integer $s\geq
0$. This spectrum is known to be organized by the $HS$ symmetry
\cite{Eastwood:2002su}, \cite{Vasiliev:2003ev} that extends $AdS$
isometries. One way to reach this spectrum is from the tensor
product of two singletons in $d-1$ dimensions, the statement known
as the Flato-Fronsdal theorem for $d=4$ \cite{Flato:1978qz},
generalized to any $d$ in \cite{Vasiliev:2004cm}. For a
comprehensive introduction into the representation theory of
singletons we refer to \cite{Bekaert:2011js}.

An observation that the $AdS$ Kerr solution somehow encodes $HS$
symmetry and has something to do with singletons can be regarded
as a yet another miraculous property of a black hole.

On a different note we remark that while single and double copies
are known to satisfy the background field equations and at the
same time the Kerr covariant ones, this is not so with the zeroth
copy and higher copies with $s>2$. We explicitly find the
'interaction' term that should be added to the black hole
covariant field equations to make them valid. Interestingly, this
term vanishes for the $s=1$ and $s=2$ cases only in any dimension.

The paper is organized as follows. In section \ref{Fronsdal} we
review the Fronsdal gauge fields. In section \ref{BH} the $AdS$
Kerr solution is given in the Kerr-Schild form along with its
double copy structure. In section \ref{mcopy} we present the
Kerr-Schild multicopy and propose a new identity for the Kerr
solution and then conclude in section \ref{concl}. The paper is
supplemented with one Appendix.

\section{Fronsdal fields}\label{Fronsdal}
Historically, the metric-like description of free spin $s$ gauge
fields was proposed by Fronsdal \cite{Fronsdal:1978rb}. The idea
was to write down the most general theory for free symmetric
fields on the Minkowski space that would be gauge invariant. The
action turns out to be fixed unambiguously
\begin{multline}
S =\\
-\half\int_{M^{d}}\Big(\p_\gm \vf^{\ga(s)}\p^\gm \vf_{\ga(s)} -
\ff{s(s-1)}{2}\p_\gm {\vf^\gn_{\gn}}^{\ga(s-2)} \p^\gm
{\vf_{\gr}^\gr}_{\ga(s-2)}+s(s-1)
\p^\gm {\vf^\gn_\gn}^{\ga(s-2)}\p^\gr \vf_{\gr \gm \ga(s-2)} -\\
-s\p_\gm \vf^{\gm \ga(s-1)}\p^\gn \vf_{\gn \ga(s-1)}-
\ff{s(s-1)(s-2)}{4}\p_\gm {\vf^\gn_\gn}^{\gm
\ga(s-3)}\p^\gr{\vf_\gr^\gc}_{\gc \ga(s-3)}\Big)\,,
\end{multline}
where we use the convention that assigns symmetrization over a
group of indices denoted by a single letter, e.g.,
\be
A^{\al}B^{\al}:=A^{\al_1}B^{\al_2}+A^{\al_2}B^{\al_1}\,,
\ee
(see Appendix B of \cite{Didenko:2014dwa}) . Varying the action,
one can obtain the dynamical equations
\be\label{eq1}
\square \vf^{\ga(s)} - \p^\ga \p_\gm \vf^{\gm \ga(s-1)}+\p^\ga
\p^\ga {\vf^{\ga(s-2)\gm}}_\gm = 0\,,
\ee
which are invariant under gauge transformations
\be
\gd \vf^{\ga(s)} = \p^\ga \xi^{\ga(s-1)} \q
\gx^{\ga(s-3)\gm}{}_\gm = 0\,.
\ee
It is not hard to recognize equations of motion for spin $s = 0$
massless scalar, Maxwell equations for $s  = 1$ and the linearized
Einstein equations for $s = 2$ field. A massless spin $s$ field is
described by a totally symmetric rank-$s$ field $\vf^{\ga(s)} =
\vf^{\ga_1\ldots\ga_s}$, which fulfills the double traceless
condition
\be\label{tr}
\vf^{\ga(s-4)\gm \gn}{}_{\gm \gn} =\vf^{\ga(s-4)\ga \gm \gb
\gn}\gpp_{\ga \gm}\gpp_{\gb \gn}= 0\,.
\ee
A brief analysis tells us that the Fronsdal equations carry a
spin-$s$ representation of the Poincare group. In addition, the
solution forms a representation of Wigner's little algebra
$so(d-2)$. Details can be found, for example, in
\cite{Ponomarev:2022vjb}.

The Fronsdal theory is developed on the maximally symmetric
backgrounds as well. Let us confine ourselves to the case of
$AdS_d$ in what follows. It corresponds to the negative
cosmological constant $\lambda<0$ with the isometry algebra being
$so(d-1,2)$. One then just replaces the Minkowski metric with the
$AdS$ one in \eqref{tr} so that, conditions on the field remain
the same
\be
\vf^{\ga(s-4)\gm \gn}{}_{\gm \gn} =\vf^{\ga(s-4)\ga \gm \gb
\gn}\bar{g}_{\ga \gm}\bar{g}_{\gb \gn}= 0\,.
\ee
The Riemann tensor for maximally symmetric space is defined as
follows
\be\label{com}
[\n_{\gm} \n_{\gn}]V^\ga = \gl \gd^\ga_\gm \overline{g}_{\gn
\gb}V^\gb -  \gl \gd^\alpha_\gn \g_{\gm \gb}V^\gb\,.
\ee
From now on we use barred derivatives $\overline{\nabla}$ for
$AdS$. Gauge transformations are modified as follows
\be
\gd \vf^{\ga(s)} = \n^{\ga}{\xi^{\ga(s-1)}}\,.
\ee
Likewise, the Fronsdal equations acquire the $AdS$ covariant form
\begin{align}\label{eq2}
& \d \vf^{\ga(s)} - \n^{\ga}\n_{\gm} \vf^{\gm\ga(s-1)} + \half
\n^{\ga}\n^{\ga} {\vf^{\ga(s-2)\gm}}_\gm - m^2_{s}\vf^{\ga(s)}
+2\lambda \g^{\ga \ga}{\vf^{\ga(s-2)\gm}}_\gm = 0\,, \\
&m^2_{s} = - \lambda((s-2)(d+s-3)-s)\,.\label{mass}
\end{align}
Note that the mass-like term \eqref{mass} is not equal to zero for
massless fields, nevertheless the equation remains gauge invariant
for $s\geq 1$. The scalar $s=0$ is not a gauge field and should be
excluded from the Fronsdal analysis, or to put it differently, its
mass is left unspecified. The analytic continuation to $s=0$ in
$m^2_{s}$, however, gives a nonzero value that is precisely the
correct mass of a scalar in $HS$ gauge theory of symmetric fields
\cite{Vasiliev:2003ev}. In the sequel we will use \eqref{eq2} for
all integer $s\geq 0$.

\section{$(A)dS$ Kerr-Schild black holes}\label{BH}
To proceed with the multicopy construction, let us consider the
$AdS$ rotating Kerr black hole solution in $d$ dimensions
originally found in \cite{Gibbons:2004js}. Its metric has the
Kerr-Schild form with the $AdS$ base metric $\g_{\mu \nu}$
\be\label{eq0}
g_{\mu \nu} = \g_{\mu \nu} + M \vf_{\mu \nu}, \qquad \vf_{\mu \nu}
= k_{\mu}k_{\nu}\vf\,,
\ee
where $\vf$ is a scalar function and $k^{\gm}$ is a null and
geodesic vector with respect to both the base metric and $g_{\gm
\gn}$
\be\label{eq}
k_\mu k^\mu = k^\nu \nabla_\nu k^\alpha = k^\nu
\overline{\nabla}_\nu k^\alpha = 0\,.
\ee
Obviously, the inverse metric is given by
\be
g^{\gm \gn} = \g^{\gm \gn} - M \vf^{\gm \gn}\,.
\ee
A detailed analysis of this metric leads to an interesting
property. Its Ricci tensor in mixed components $R^\gm_{\gn}$ is
linear in $\vf$
\be\label{eq3}
R^{\gm}_{\gn} = \overline{R}^{\gm}_{\gn} -
\vf^{\gm}_{\rho}\overline{R}^{\rho}_{\gn} + \half\n_\rho \n_\gn
\vf^{\gm \rho} + \half\n^\rho \n^\gm \vf_{\gn \rho} - \half\n^\rho
\n_\rho \vf^\gm_\gn\,,
\ee
where indices of barred objects are raised and lowered by the base
metric $\g_{\mu \nu}$. To give an explicit coordinate realization
let us, following \cite{Gibbons:2004js}, introduce the $(A)dS_d$
metric in spheroidical coordinates. The realization is different
for odd and even $d$. Consider the case of $d = 2n$
\begin{multline}
    \overline{g}_{\mu \nu}dx^{\mu}dx^{\nu} = -W(1-\lambda r^2)dt^2 + Fdr^2 +\sum^n_{i = 1}\frac{(r^2 +a^2_i)}{1 +\lambda a^2_i}d\mu_i^2 + \sum^{n-1}_{i = 1}\frac{(r^2 +a^2_i)}{1 +\lambda a^2_i}\mu_i^2 d\phi_i^2 +\\
    +\frac{\lambda}{W(1-\lambda r^2)}\left(\sum^n_{i=1}\frac{(r^2+a_i^2)\mu_i d\mu_i}{1 +\lambda
    a^2_i}\right)^2\,,
\end{multline}
where $a_i$ are free parameters. The corresponding null vector and
scalar function that form a black hole solution are
\begin{equation}\label{coord}
k_{\mu}dx^{\mu} = Wdt + Fdr - \sum_{i = 1}^{n-1}\frac{a_i
\mu^2_i}{1+\lambda a_i^2}d\phi_i,  \qquad \vf = \frac{1}{r\sum_{i
= 1}^{n}\frac{\mu^2_i}{r^2+a_i^2}\prod_{j = 1}^{n-1}(r^2 +
a_j^2)}\,,
\end{equation}
where $\phi_i$ are angular coordinates and
\begin{equation}
W \equiv \sum_{i = 1}^{n}\frac{\mu^2_i}{1+\lambda a_i^2}, \qquad F
\equiv \frac{r^2}{1-\lambda r^2}\sum_{i =
1}^{n}\frac{\mu^2_i}{r^2+a_i^2}\,.
\end{equation}
Coordinates (momenta) $\gm_i$ are subject to the constraint
\begin{equation}
\sum_{i = 1}^{[d/2]} \gm^2_i = 1\,.
\end{equation}
For the odd case $d = 2n+1$ there is an analogous formula (see
\cite{Gibbons:2004js}).

\paragraph{Black hole as a double copy}
Now we are ready to construct a zeroth copy out of the Kerr-Schild
data. Specifically, it can be checked
\cite{Carrillo-Gonzalez:2017iyj} that $\vf$ does satisfy the
equation\footnote{The notation of \cite{Carrillo-Gonzalez:2017iyj}
is related to ours as follows $\overline R=d(d-1)\gl$.}
\begin{equation}\label{eq4}
\left(\overline{\Box} -2\gl\,(d-3)\right)\vf = 0\,.
\end{equation}
Following \cite{Carrillo-Gonzalez:2017iyj} we have also checked
the mass-like term above using {\it Mathematica} in $4\leq d \leq
12$ dimensions. While we do not have proof of the validity of
\eqref{eq4} in all dimensions, we find it is satisfied for all
$4\leq d\leq 12$. We believe it holds in any $d$.

Similarly, the single copy defined as $\vf^{\mu} = \vf k^{\mu}$
satisfies Maxwell's equations on $(A)dS$ background
\begin{equation}\label{fr1}
\n_{\mu}F^{\mu \nu} = \n_{\mu} (\n^{\mu} \vf^{\nu} - \n^{\nu}
\vf^{\mu}) = 0\,.
\end{equation}
In order to complete our system, we should write out an equation
for double copy $\vf^{\gm \gn} = \vf k^{\gm} k^{\gn}$, which is
the solution of the linearized (and exact) Einstein's equations.
To do that, we first change the order of covariant derivatives in
\eqref{eq3} using \eqref{com}.
In our case,
\be
[\n^\alpha \n_{\mu}]\vf^{\mu\beta} = -\lambda d\vf^{\alpha
\beta}\,.
\ee
Second, we remember that metric $\g_{\gm \gn}$ satisfies
$\overline{R}_{\gm \gn} = \gl (d-1)\g_{\gm \gn}$, then the full
metric satisfies Einstein's equations   $R_{\gm \gn} = \gl
(d-1)g_{\gm \gn}$ with the same cosmological constant, so that
\eqref{eq3} becomes
\be\label{fr2}
\overline\Box \vf^{\gm \gn} - \n^\gm\n_\gr \vf^{\gr \nu} -
\n^\gn\n_\gr \vf^{\gr \gm} - 2\gl \vf^{\gm \gn} =0\,.
\ee
One can recognize in \eqref{eq4}, \eqref{fr1} and \eqref{fr2} the
particular cases of the Fronsdal equations for spins $s = 0$, $s =
1$, and $s = 2$, respectively.

\section{Fronsdal multicopy solutions}\label{mcopy}
The natural extension of the Kerr-Schild double copy for arbitrary
spin field is \cite{Didenko:2008va}
\be\label{eq5}
\vf^{\ga(s)} = k^{\ga_1}\ldots k^{\ga_s} \vf
\ee
where we have $s$ copies of vector $k^{\gl}$. Our goal is to write
out an equation for this field and the statement is that it solves
the Fronsdal equations \eqref{eq2} for any $s\geq 0$. To prove
this we use the system of equations for zeroth, single, and double
copies introduced above
\be\label{eq6}
\left\lbrace
\begin{aligned}
&\left(\overline\Box- 2(d-3)\gl\right)\vf = 0 \,, \\
&\overline\Box(\vf k^\gm) - \n_\gn \n^\gm(\vf k^{\gn}) = \overline\Box(\vf k^\gm) - \n^\gm \n_\gn(\vf k^{\gn}) -\gl(d-1)\vf^{\gm} =0 \,,\\
&\overline\Box \vf^{\gm \gn} - \n^\gm\n_\gr \vf^{\gr \gn} -
\n^\gn\n_\gr \vf^{\gr \gm} - 2\lambda \vf^{\gm \gn} =0
\end{aligned}\right.
\ee
and the commutation relation for covariant derivatives
  \begin{equation}
[\n^\alpha \n_{\mu}]\vf^{\mu \ga(s-1)} = -\lambda
(d+s-2)\vf^{\ga(s)}\,.
 \end{equation}
Decomposing the d'Alembert operator and taking into account the
system above, results in
\be\label{eq7}
\d \vf^{\ga(s)} = \d\vf^{\ga(s-1)}k^{\ga_s} + 2\n_\rho
\vf^{\ga(s-1)} \n^\rho k^{\ga_s} + \vf^{\ga(s-1)} \d k^{\ga_s}
=\n^{(\ga_s}\n_m \vf^{m\ga(s-1))} + m^2_{s} \vf^{\ga(s)},
\ee
leading eventually to
\begin{align}\label{fr}
&\d \vf^{\ga(s)} - \n^{\ga}\n_{\gm} \vf^{\gm\ga(s-1)}  -
m^2_{s}\vf^{\ga(s)} = 0\,,
\end{align}
where $m^2_{s}$ is given by \eqref{mass}. Note that since
\be
\vf^{\al(s-2)\gb}{}_{\gb}=0
\ee
due to \eqref{eq}, the traceful terms vanish. This implies that
from \eqref{fr} multicopy \eqref{eq5} satisfies the Fronsdal
equations \eqref{eq2} for all $s\geq 0$. The details of the
derivation of \eqref{fr} are given in the Appendix.

A few comments are now in order. The Kerr-Schild double copy sets
the mapping of fields $s = 0,1,2$ to higher-spin ones. Indeed, eq.
\eqref{fr} for $s>2$ is a consequence of the lower-spin copies
\eqref{eq6} and Kerr-Schild condition \eqref{eq}. In particular,
the mass-like term \eqref{mass} which turns out to be exactly the
one of the Fronsdal theory originates from the very specific
zeroth copy mass in \eqref{eq4}. This value is not accidental as
it appears to be the one that comes from tensor product of two
$AdS$ scalar singletons. This fact is known in four dimensions as
the Flato-Fronsdal theorem \cite{Flato:1978qz} generalized to any
$d$ in \cite{Vasiliev:2004cm}. From that perspective it is not
surprising that \eqref{eq6} gives rise to all symmetric massless
fields satisfying \eqref{eq2} as multicopies. Less clear is why
the $AdS$ Kerr rotating solution generates exactly this particular
zeroth copy mass. In four dimensions there is a neat derivation of
the Weyl double and multicopies that highlights its close relation
to massless (conformal) fields in $AdS_4$ based on the Penrose
transform \cite{Didenko:2009td}, \cite{White:2020sfn},
\cite{Didenko:2021vui}. For arbitrary $d$ we are not aware of any
similar explanation as $m^2_{0}$ no longer corresponds to the
conformal case in general.

Another interesting fact that reveals a link between the
Kerr-Schild scalar and vector field is observed in various
dimensions using {\it Mathematica}. Namely, one can check the
following relation
\be\label{id}
\nn^{\gm_1}\nn^{\gm_2}\ldots \nn^{\gm_{d-3}}(k_{\gm_1}\ldots
k_{\gm_{d-3}}) =(d-2)! \vf\,,
\ee
where $\nn$ can be equally well replaced with $\overline\nn$. The
relation between $\vf$ and $k^{\mu}$ involves higher derivatives,
the number of which grows linearly with space-time dimension $d$.
It says, in particular, that function $\vf$ is a scalar with
respect to either the black-hole or $AdS$ metric. Indeed,
substituting \eqref{id} into the Kerr-Schild ansatz \eqref{eq0} in
place of $\vf$ one notices that the metric properly transforms
under diffeomorphisms provided $k^{\mu}$ transforms as a vector
and therefore $\vf$ is a true scalar. This in turn implies that
the particular coordinate realization \eqref{coord} of the
Kerr-Schild ansatz is not important.

\paragraph{$\nn$-covariant form of multicopy}

A curious fact about single and double copies that satisfy
\eqref{fr} for $s=1$ and $s=2$ is that the background derivative
$\overline\nn$ in \eqref{fr} can be equivalently replaced with the
Kerr-Schild one $\nn$ such that it does not spoil solutions
\be\label{s1cov}
\overline\Box(\vf k_\gm) - \n_\gn \n_\gm(\vf k^{\gn}) =\Box (\vf
k_\gm) - \nn_{\gn} \nn_{\gm}(\vf k^{\gn}) = 0\,,
\ee
\begin{multline}\label{s2cov}
\overline\Box\vf^{\gm \gn} - \n_\gr \n^{\gm}\vf^{\gr \gn} - \n_\gr \n^{\gn} \vf^{\gr \gm}+\gl(d-1)\vf^{\gm \gn} = \\
\Box\vf^{\gm \gn} - \nn_\gr \nn^{\gm}\vf^{\gr \gn} - \nn_\gr
\nn^{\gn} \vf^{\gr \gm} +\gl(d-1)\vf^{\gm \gn} = 0\,.
\end{multline}
The reason is purely kinematical and rests on Kerr-Schild
conditions \eqref{eq}. This is not the case, however, with the
zeroth copy $s=0$, $\vf$ which enjoys \eqref{eq4} in the $AdS$
background only
\be
(\overline{\square} - 2\gl (d-3)) \vf \neq  ({\square} - 2\gl
(d-3)) \vf\,.
\ee
It is therefore clear that for the full Fronsdal system \eqref{fr}
one can not replace background derivatives with the Kerr ones
without any effect. Still, in doing so one may ask what kind of
terms one should add to compensate such a replacement? To this end
consider zeroth copy $\vf$. Using \eqref{eq4} and \eqref{eq} we
derive
\be\label{s0cov}
\Box\vf-m_{0}^2\vf+M\nn_{\gga}\left(\vf^{\gb\gga}\nn_{\gb}\vf\right)=0\,.
\ee
Note, that the last term on the right \eqref{s0cov} is
proportional to the black-hole mass parameter $M$. Using
\eqref{s1cov}, \eqref{s2cov}, \eqref{s0cov} and \eqref{eq} it is
not difficult to come up with the following $\nn$-covariant form
of equations for multicopy $\vf^{\al(s)}$
\begin{multline}\label{eq8}
\square \vf^{\ga_1 \ldots \ga_s} - \nn_{\gga}\nn^{(\ga_1}
\vf^{\ga_2 \ldots \ga_s)\gga}  - m^2_{s}\vf^{\ga_1 \ldots \ga_s}
+M\frac{(s-1)(s-2)}{2}\nn_\gga [\vf^{\beta \gga}\nn_\beta
\vf^{\ga_1 \dots \ga_s}]=0\,.
\end{multline}
Note that the last term on the left of \eqref{eq8} vanishes for
$s=1$ and $s=2$ and is never zero for the rest. The correction
proportional to $M$ comes in the form of a total derivative. Up to
a normalization this result is in agreement with the $d=4$ case
considered earlier in \cite{Didenko:2008va}.

Let us also stress that the Fronsdal equations \eqref{eq2} do not
admit covariantization to an arbitrary background. The naive
replacement of the $(A)dS$ covariant derivatives with those from
less symmetric geometry would result in a loss of gauge invariance
due to $[\nn, \nn]\sim$ Riemann and correspondingly to the
appearance of extra degrees of freedom. From that perspective the
presence of the 'interaction' term in \eqref{eq8} comes as no
surprise. It would be interesting to analyze \eqref{eq8} relying
on explicit cubic $HS$ interactions using the Kerr-Schild
formalism.

\section{Conclusion}\label{concl}

In this letter we extend the known results on classical
Kerr-Schild double copy to symmetric higher spins in any
dimensions on $(A)dS$ background. Similarly to the standard case,
where double copy results from squaring a single copy, the
higher-spin $s$ multicopy appears as power $s$ of a single copy up
to the zeroth one. Such an extension goes naturally for all
integer spins $s\geq 0$ introducing higher spins $s>2$ on equal
footing with the lower ones. The copies respectively satisfy the
Klein-Gordon, Maxwell and Einstein equations and their analogs for
$s>2$ the Fronsdal equations \eqref{eq2}.

Interestingly, while the zeroth copy corresponds to a conformal
scalar in $d=4$, for general $d$ its 'mass' is no longer conformal
\cite{Carrillo-Gonzalez:2017iyj} though is fixed in terms of
space-time dimension and the cosmological constant \eqref{m0}. To
the best of our knowledge its particular value seems to have no
relevant explanation beyond\footnote{The case of $d=6$ corresponds
to conformal coupling of the zeroth copy too. However unlike
$d=4$, none of the single, double or any multicopies carry
representations of conformal algebra in that case. We thank the
anonymous Referee for pointing this out to us.} $d=4$ in the
double copy literature. We remark that this value is precisely the
one that results from higher spin symmetry in any dimension (see
also \cite{Bekaert:2011js} for a nice introduction of the
singleton point of view). Along with all symmetric gauge fields
$s\geq 0$, it gives field spectrum \eqref{spect}, where the
multicopy naturally shows up via \eqref{anys}. Given higher-spin
symmetry plays a fundamental role in field theories in the
unbroken phase \cite{Vasiliev:2014vwa}, one may argue that the
double copy philosophy should be considered more broadly then just
relations between gravity and gauge theory quantities.

On the other hand, looking at Kerr solution \eqref{KS}, we are
curious to know how and why a black hole 'knows' about higher-spin
symmetry. Indeed, scalar $\vf$ entering metric \eqref{KS} for some
reason results in the mass of the zeroth copy that matches exactly
the one that comes from tensor product of two Dirac singletons.
Had it been different, there would be no Fronsdal gauge fields as
multicopies. Whether this remarkable property comes from the
generalized type $D$ of $d$-dimensional Kerr solution
\cite{Coley:2004jv} or has deeper grounds remains unclear. It
would be very interesting to trace the relation of the Kerr-Schild
ansatz to the Flato-Fronsdal theorem \cite{Flato:1978qz},
\cite{Vasiliev:2004cm} from the representation theory standpoint.
For an interesting proposal in that direction we refer to
\cite{Iazeolla:2008ix}, \cite{Iazeolla:2020jee}. At the same time,
one can arrive at seemingly any zeroth copy profile from the
nonlinear electrodynamics along the lines of
\cite{Mkrtchyan:2022ulc} thus breaking down the Fronsdal
multicopy. In this sense the Kerr black hole can be viewed as an
example of a gravitational solution corresponding to unbroken $HS$
symmetries.

Let us also point out a new interesting identity \eqref{id} that
relates higher derivatives of the Kerr-Schild vectors to zeroth
copy $\vf$. This identity in particular allows one to get rid of
the scalar function $\vf$ from the Kerr-Schild ansatz rewriting
the latter in terms of the Kerr-Schild vector $k^\mu$ only.

In the case of $d=4$, for example, the appearance of gauge fields
via multicopies was crucial in generalizing the Kerr solution into
nonlinear higher-spin theory \cite{Didenko:2009td}, (see also
\cite{Iazeolla:2011cb} for generalization). In a different context
the multicopy based solutions seem to play an important role in
$HS$ holography \cite{David:2020fea}-\cite{Lysov:2022nsv}.

In conclusion, the results of this paper can be viewed as a
$d$-dimensional generalization of the earlier analysis of four
\cite{Didenko:2008va} and five \cite{Didenko:2011ir} dimensions,
where a spinor isomorphism could play a distinguished role.
Indeed, in $d=4$ the related Weyl multicopy has a transparent
origin as the Penrose transform \cite{Didenko:2009td},
\cite{Neiman:2017mel}, \cite{White:2020sfn},
\cite{Didenko:2021vui}, which naturally generates the whole tower
of massless fields of any spin. In particular, the zeroth copy
mass is exactly the one of a conformal scalar in that case. Our
case of general dimensions seemingly lacks such an interpretation.

\section*{Acknowledgments}
We are grateful to Mitya Ponomarev for valuable comments on the
draft of the paper. VD would like to thank Carlo Iazeolla for
stimulating discussion on black holes in the context of the
Flato-Fronsdal theorem and Mariana Carrillo-Gonz\'{a}lez for
correspondence. The research was supported by the RFBR grant No
20-02-00208.

\newcounter{appendix}
\setcounter{appendix}{1}
\renewcommand{\theequation}{\Alph{appendix}.\arabic{equation}}
\addtocounter{section}{1} \setcounter{equation}{0}
\renewcommand{\thesection}{\Alph{appendix}.}

\section*{Appendix. Checking solutions of Fronsdal equations}\label{Appendix:app1}
Let us consider fields \eqref{eq5}. Using the Leibniz rule, we
rewrite d'Alembertian as follows
\be\label{eq14}
\left\lbrace
\begin{aligned}
&\d \vf^{\ga} = \d \vf k^\ga + 2\n_{\rho}\vf\n^{\rho}k^\ga + \vf \d k^\ga\,,\\
&\d \vf^{\ga\gb} = \d \vf^{\ga} k^\gb +
2\n_{\rho}\vf^{\ga}\n^{\rho}k^\gb + \vf^{\ga} \d k^\gb\,.
\end{aligned}\right.
\ee
Multiplying the first equation \eqref{eq14} with $k^{\al}$ and
subtracting it from the second
\begin{equation}
2\vf \n_\gd k^\ga \n^\gd k^\gb = \d \vf^{\ga\gb} - k^\ga \d
\vf^\gb - k^\gb \d \vf^\ga + k^\ga k^\gb \d \vf = \n^\ga k^\gb
\n_\gd \vf^\gd + \n^\gb k^\ga \n_\gd \vf^\gd - 2\lambda
\vf^{\ga\gb}\,.
\end{equation}
Using the light-like and geodesic conditions on vectors $k^\al$,
\eqref{eq3} we have
\begin{multline}\label{eq15}
2\vf \n_\gd (k^\ga k^\gb) \n^\gd k^\gga =\n^\gb k^\gga \n_\gd
\vf^{\gd\ga} + \n^\ga k^\gga \n_\gd \vf^{\gd\gb} + \n^\gga( k^\ga
k^\gb) \n_\gd \vf^\gd - 4\lambda \vf^{\ga\gb \gga}\,.
\end{multline}
Now, let us write out the action of the d'Alembert operator on the
spin $s=3$ field upon substituting ($\ref{eq6}$), ($\ref{eq14}$),
($\ref{eq15}$)
\begin{multline}
\d(\vf^{\ga\gb\gga}) =\d \vf^{\ga\gb} k^\gga + 2\n_{\gd}\vf^{\ga\gb}\n^{\gd}k^\gga + \vf^{\ga\gb} \d k^\gga =(\n^\ga \n_\gd \vf^{\gd\gb} + \n^\gb \n_\gd \vf^{\gd\ga}+2\lambda \vf^{\ga\gb})k^\gga + 2\n_{\gd}\vf^{\ga\gb}\n^{\gd}k^\gga + \\
+k^\ga k^\gb(\n^\gga \n_\gd \vf^{\gd} + \lambda(d-1)\vf^\gga - 2\lambda k^\gga(d-3) - 2\n_{\gd}\vf\n^{\gd}k^\gga )\\ = k^\gga \n^\ga\n_{\gd} \vf^{\gd\gb} + k^\gga \n^\gb \n_{\gd}\vf^{\gd\ga} + k^\ga k^\gb \n^\gga \n_{\gd} \vf^{\gd} - \lambda(d-7)\vf^{\ga\gb\gga}
+2\vf\n_{\gd}(k^\ga k^\gb)\n^{\gd}k^\gga = \\
=\n^\ga(k^\gga k^\gb \n_\gd \vf^\gd) + \n^\gb(k^\ga k^\gga \n_\gd \vf^\gd) + \n^\gga(k^\ga k^\gb \n_\gd \vf^\gd) -\lambda(d-3)\vf^{\ga \gb \gga}  =\\
=\n^\ga \n_\gd \vf^{\gd\gb\gga} + \n^\gb\n_\gd \vf^{\gd\ga\gga} +
\n^\gga \n_\gd \vf^{\gd\ga\gb} - \lambda(d-3)\vf^{\ga \gb \gga}\,.
\end{multline}
We observe here that the Fronsdal equations for spin $s=3$ field
\eqref{eq2} are satisfied. Now by induction, assume that Fronsdal
equations \eqref{eq2} are satisfied for the Kerr-Shield fields
with the spin $s-1$
\begin{equation}\label{eq16}
\d \vf^{\ga(s-1)}- \n^\ga \n_\gm \vf^{\gm \ga(s-2)} +
\lambda((s-3)(d+s-4)-s+1)\vf^{\ga(s-1)} = 0\,.
\end{equation}
Eq. ($\ref{eq15}$) can be generalized to the following one
\begin{equation}\label{eq17}
2\vf \n_\gm (k^\ga\ldots k^{\ga_{s-1}}) \n^\gm k^{\ga_s} = \n_\gm \vf^{\gm(\ga(s-2)}\n^{\ga_{s-1})}k^{\ga_{s}}+\n_\gm \vf^{\gm}\n^{\ga_s}(k^{\ga_1}\ldots k^{\ga_{s-1}}) - 2\lambda(s-1)\vf^{\ga(s)}
\end{equation}
where $(\ldots)$ is the symmetrization over indices. Writing up
the d'Alembert operator for the spin $s$ field and substituting
\eqref{eq17}, \eqref{eq16}, \eqref{eq14}, \eqref{eq6} we finally
obtain
\begin{multline}\label{eq18}
\d \vf^{\ga(s)} = \d\vf^{\ga(s-1)}k^{\ga_s} + 2\n_\gm \vf^{\ga(s-1)} \n^\gm k^{\ga_s} + \vf^{\ga(s-1)} \d k^{\ga_s} = \\
k^{\ga_s}\n^{\ga_{s-1}} \n_\gm \vf^{\gm \ga(s-2)}-\gl((s-3)(d+s-4)-s+1)\vf^{\ga(s)} + \n_\gm \vf^{\gm(\ga(s-2)}\n^{{\ga_{s-1}})}k^{\ga_{s}}+\\
+\n_\gm \vf^{\gm}\n^{\ga_s}k^{\ga(s-1)} -2\gl(s-1)\vf^{\ga(s)}  + k^{\ga(s-1)}\underbrace{(\gl(d-1)\vf^{\ga_s}+\n^{\ga_s}\n_\gm \vf^\gm - 2\gl(d-3)\vf^{\ga_s})}_{(s = 1) - (s = 0)}=
\\k^{\ga_s}\n^{\ga_{s-1}} \n_\gm \vf^{\gm \ga(s-2)}  + \n_\gm \vf^{\gm(\ga(s-2)}\n^{\ga_{s-1})}k^{\ga_{s}} + k^{\ga(s-1)}\n^{\ga_s}\n_\gm \vf^\gm + \n_\gm \vf^{\gm}\n^{\ga_s}k^{\ga(s-1)}-\\ -\gl((s-2)(d+s-3)-s)=
\n^{(\ga_{s-1}}(k^{\ga_s}\n_\gm \vf^{\gm\ga(s-2))})+\n^{\ga_{s}}(k^{\ga(s-1)}\n_\gm \vf^{\gm})+m^2\vf^{\ga(s)}=
\\
=\n^{(\ga_s}\n_\gm \vf^{\gm\ga(s-1))} + m^2 \vf^{\ga(s)}\,,
\end{multline}
where $k^{\ga(s)} = k^{\ga_1}\ldots k^{\ga_s}$. Now one observes
that Fronsdal equations \eqref{eq2} do satisfy.

\end{document}